\documentclass[reprint,aps,prb]{revtex4-2}
\usepackage{dcolumn}
\usepackage{braket}%braketのコマンド
\usepackage{amsmath}
\usepackage{amssymb}
\usepackage{mathtools}
\usepackage[dvipdfmx]{graphicx}
\usepackage{bm}

\begin{document}

\title{Enhanced valley polarization of graphene on hBN under circularly polarized light irradiation}

\author{Keisuke Nakagahara$^1$}
\author{Katsunori Wakabayashi$^{1,2,3}$}
\affiliation{$^1$Department of Nanotechnology for Sustainable Energy, School of Science and Technology, Kwansei Gakuin University, Gakuen-Uegahara 1, Sanda 669-1330, Japan}
\affiliation{$^2$National Institute for Materials Science (NIMS), Namiki 1-1, Tsukuba 305-0044, Japan}
\affiliation{$^3$Center for Spintronics Research Network (CSRN), Osaka
University, Toyonaka 560-8531, Japan}

\begin{abstract}
  Graphene on hBN (G/hBN) has a long period moir\'{e} superstructure owing to
 the lattice mismatch between two materials. Long periodic potential
 caused by the moir\'{e} superstructure induces modulation of electronic
 properties of the system. In this paper, we numerically calculate
 optical conductivity of G/hBN under circularly polarized light
 irradiation. The lack of spatial inversion symmetry in G/hBN induces
 the valley polarization. In further, the valley polarization becomes
 most pronounced in the infrared and terahertz regions if the twist angle between two materials is close to
 zero for non-doping case, however, insensitive with twist angle for
 hole-doped case. These results will serve to design the valleytronics
 devices using G/hBN.
\end{abstract}

\maketitle

%\section{Introduction}
%After the successful isolation of
%graphene~\cite{novoselov2004electric}, 
%two-dimensional material are attracting much attention.

Two-dimensional (2D) atomically-thin materials are attracting much
attention owing to its high flexibility to control the electronic and
optical properties to design new functional devices\cite{kostya2016science,das2015,Schaibley2016,Zheng2018rev}.
Graphene, one atomic thickness carbon sheet, is one of the fundamental 2D materials\cite{RevModPhys.81.109}.
Since graphene has honeycomb lattice structure,
and its electronic states near the Fermi energy are well described by 
massless Dirac equation\cite{ajiki1993,ando1998}.
When two graphenes are overlaid with slight twisting angle, 
the moir\'{e} superstructure appears and works as the 
long periodic electronic potential for Dirac electrons of graphene
\cite{dos2007graphene,trambly2010localization,mele2010commensuration,bistritzer2011moire,morell2010flat,moon2012energy,dos2012continuum,moon2013optical}, 
i.e., so-called {\it twisted bilayer graphene} (TBG). 
The period of the moir\'{e} superstructure becomes larger with decreasing the
twist angle.
TBG provides attractive properties such as vanishing the Fermi
velocity\cite{bistritzer2011moire}, flat band at the magic
angle\cite{Utama2021.natphys,Lisi2021.natphys,sato2021} and superconductivity\cite{cao2018correlated,cao2018unconventional,yankowitz2019tuning}. 
In addition to TBGs, moir\'{e} superstructure also appears when different
materials are overlaid.
Graphene on hBN (G/hBN), i.e., graphene overlaid on hexagonal boron nitride (hBN),
has moir\'{e} superstructure owing to the mismatch of lattice constants between two materials\cite{PhysRevLett.115.186801,PhysRevLett.113.135504,Yankowitz2019.natrevphys}.
Experiments of G/hBN have revealed many intriguing phenomena such as
the Hofstadter butterfly\cite{dean2013hofstadter,hunt2013massive}
and the fractional quantum Hall effect \cite{Wang2015.science,Yankowitz2019.natrevphys,10.1038/s41567-020-0906-9}.

Meanwhile, 2D materials with honeycomb lattice structure such as graphene
and hBN have local minima in the conduction band and local maxima in the
valence band in the momentum space, which are referred as {\it valley}. 
Recently much efforts are devoted to manipulate the valley degree of
freedom to encode and process information, i.e., {\it valleytronics},
which is valley analogue of spintronics\cite{Schaibley2016,Rycerz2007,Xiao2012.prl,Xu2014.natphys,Shkolnikov2002,Isberg2013,Zhu2012.naturephys,Shimazaki2015.natphys}.
The irradiation of circular polarized light onto 2D materials is one of
representative ways to generate the valley polarized
states\cite{yao2008valley,cao2012valley,zeng2012valley,Xiao2012.prl,mak2012control,Jones2013.natnanotech}.  
Since the broken spatial inversion symmetry is needed to induce
valley polarization, 
the valley polarized states can not be realized in monolayer graphene which respects the
spatial inversion symmetry. 
However, the moir\'{e} superlattice potential of G/hBN 
owing to the lattice mismatch has the trigonal symmetry, i.e., no inversion symmetry,  
resulting in the valley polarization.
Several groups have recently investigated the electronic
properties of G/hBN\cite{wallbank2013generic,moon_ghbn,PhysRevResearch.2.043427}, and optical
absorption
properties\cite{abergel2013infrared,shi2014gate,abergel2015infrared,dasilva2015terahertz}. 
However, the optical properties under circular polarized light in G/hBN have not been investigated yet.
In this paper, we numerically calculate optical conductivity under
circular polarized light irradiation in G/hBN heterostructure.
It is shown that the valley polarization is induced in G/hBN owing to
the lack of spatial inversion symmetry. In further, the valley
polarization becomes most pronounced when the twist angle between
graphene and hBN is close to zero for non-doping case.
However, for hole-doped case, the valley polarization is insensitive to 
the twist angle.

Figure 1(a) shows crystal structure of G/hBN with a twist angle
$\theta=0^{\circ}$,
where moir\'{e} superstructure is clearly seen owing to the lattice mismatch.
$\bm{L}_1^{\mathrm{M}}$ and $\bm{L}_2^{\mathrm{M}}$ are moir\'{e}
primitive vectors and the area surrounded by rhombus indicates moir\'{e}
unit cell. 
The lattice constant of graphene is $a = 0.246$ nm.
Though hBN has the same honeycomb lattice
structure,
the lattice constant of hBN is 
$a_{\mathrm{hBN}} = 0.2504$ nm\cite{Liu2003, moon_ghbn},
which is slightly larger than graphene. The ratio of lattice 
constant between graphene and hBN is $\alpha = a_{\mathrm{hBN}}/a \approx 1.018$. 
When we define $\bm{a}_1 = a(1,0)$ and $\bm{a}_2 = a(1/2,\sqrt{3}/2)$ as
the primitive vectors of graphene, the primitive vectors of hBN layer of
G/hBN are
defined as $\tilde{\bm{a}} = \alpha R(\theta)\bm{a}_i \ (i=1,2)$,
where $R(\theta)$ is the rotation matrix by angle $\theta$. 
We define the reciprocal vectors $\bm{G}_i$, $\tilde{\bm{G}}_i$ for graphene and hBN,
respectively. These vectors should satisfy the relation $\bm{a}_i\cdot\bm{G}_j = \tilde{\bm{a}}_i\cdot\tilde{\bm{G}}_j=2\pi\delta_{ij}$.
\begin{figure*}[ht]
  \begin{center}
    \includegraphics[width=0.85\textwidth]{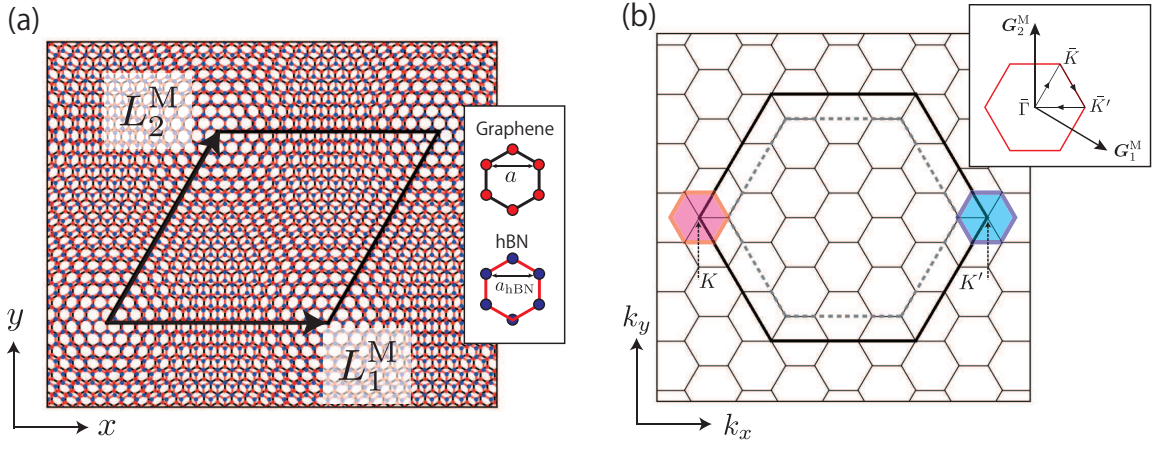}
  \end{center}
  \caption{(a) Crystal structure of G/hBN in real space for twist angle
 $\theta=0^{\circ}$. Here, $a_{\mathrm{hBN}}/a =
 16/15$ is used to draw the figure for simplicity. (b) Schematic of
 reciprocal space.
Here, $a_{\mathrm{hBN}}/a = 5/4$ is used to draw the figure for
 simplicity. 
Bold black hexagon with a thick line and gray dashed
 hexagon show the 1st BZ for graphene and hBN, respectively.
 Small hexagons are moir\'{e} BZ.
The inset is zoom of moir\'{e} BZ.
} 
 \label{fig:fig1}
\end{figure*}

The reciprocal vectors for moir\'{e} period of G/hBN can be defined as 
\begin{equation}
  \bm{G}_i^{\mathrm{M}} = (\bm{1} - \alpha^{-1}R(\theta))\bm{G}_i.
\end{equation}
Note that the period of the moir\'{e} superstructure also depends on twist angle.
The corresponding moir\'{e} primitive vectors satisfy the relation
\begin{equation}
  \bm{L}_i^{\mathrm{M}}\cdot\bm{G}_j^{\mathrm{M}} = 2\pi\delta_{ij}.
\end{equation}
Long periodic potential appears due to the moir\'{e} superstructure and affects
on the electronic states of G/hBN.
Figure 1(b) shows the schematic of reciprocal space for G/hBN.
Bold black hexagon with a thick line and gray dashed hexagon show the
1st BZ for graphene and hBN, respectively. The small hexagons are
moir\'{e} BZ spanned by $\bm{G}_1^{\mathrm{M}}$ and $\bm{G}_2^{\mathrm{M}}$.
The inset is the zoom of moir\'{e} BZ, where
$\bar{\Gamma}$ point means original $K\ (K')$ point of graphene.
In the continuum model, ${K}$ and ${K^\prime}$ valleys are
treated independently, and the energy bands can be separately
plotted in the BZ centered at ${K}$ (magenta hexagon) and that centered
at ${K^\prime}$ (cyan hexagon).

The electronic states near the Fermi energy of G/hBN are governed by the
electronic states of graphene near $K$ and $K^\prime$ points.
In this paper, we employ effective continuum model
to calculate eigenenergies and eigenfunctions of G/hBN\cite{moon_ghbn}. 
The $K$ points of graphene are located at 
$K_\xi=-\xi(2\bm{G_1}+\bm{G_2})/3$, where $\xi=\pm 1$ for $K$ and $K^\prime$, respectively.
The Dirac Hamiltonian of monolayer graphene near $K_\xi$ can be written as
\begin{equation}
H_{\mathrm{G}} = -\hbar v\bm{k\cdot\sigma_\xi},
\end{equation}
where $\bm{k}=(k_x, k_y)$ is the wave number measured from $K_\xi$ point, 
and 
$\bm{\sigma_\xi}=(\xi\sigma_x, \sigma_y)$
with Pauli matrices $\sigma_x$ and $\sigma_y$.
$H_{\mathrm{G}}$ is $2\times 2$ matrix for the basis
$\left\{A_{\xi},B_{\xi}\right\}$, i.e., sublattice degree of freedom of graphene.
The parameter $v$ is group velocity of the Dirac cone, which is given as
$v = 0.80 \times 10^6$ m/s\cite{moon_ghbn}.
By eliminating the basis for hBN using second-order perturbation theory,
the effective $2\times 2$ Hamiltonian of G/hBN is written as 
\begin{equation}
  H_{\mathrm{G/hBN}} =  H_{\mathrm{G}} + V_{\mathrm{hBN}},
  \label{eq:eff_hamil}
\end{equation}
where 
$V_{\mathrm{hBN}}$ is the long periodic moir\'{e} potential including
the effect of hBN. $V_{\mathrm{hBN}}$ is given as
\begin{eqnarray}
  \label{eq:eff_hamil}
  V_{\mathrm{hBN}} &=& V_0\left(
  \begin{array}{cc}
    1 & 0\\
    0 & 1\\
  \end{array}
  \right)\nonumber\\
  &\quad& + \left\{V_1\mathrm{e}^{i\xi\psi}\left[\left(
    \begin{array}{cc}
    1 & \omega^{-\xi}\\
    1 & \omega^{-\xi}\\
  \end{array}
    \right)\mathrm{e}^{i\xi \bm{G}_1^{\mathrm{M}} \cdot \bm{r}}\right.\right.\nonumber\\
    &\quad& + \left.\left.\left(
    \begin{array}{cc}
    1 & \omega^{\xi}\\
    \omega^{\xi} & \omega^{-\xi}\\
  \end{array}
    \right)\mathrm{e}^{i\xi \bm{G}_2^{\mathrm{M}} \cdot \bm{r}}\right.\right.\nonumber\\
    &\quad& + \left.\left.\left(
    \begin{array}{cc}
    1 & 1\\
    \omega^{-\xi} & \omega^{-\xi}\\
  \end{array}
    \right)\mathrm{e}^{i\xi \bm{G}_3^{\mathrm{M}} \cdot \bm{r}}\right]+\mathrm{H.c.}\right\},\nonumber
\end{eqnarray}
where $\bm{G}_3^{\mathrm{M}} =
-(\bm{G}_1^{\mathrm{M}}+\bm{G}_2^{\mathrm{M}})$, $v = 0.80 \times 10^6$
m/s, $V_0 = 0.0289$ eV, $V_1 = 0.0210$ eV, and $\psi = -0.29$
(rad)\cite{moon_ghbn}.
Owing to $V_{\mathrm{hBN}}$, a state at $\bm{k}$ is related to the other
states at $\bm{k}+n_1\bm{G}_1^\mathrm{M}$, $\bm{k}+n_2\bm{G}_2^\mathrm{M}$,
and $\bm{k}+n_3\bm{G}_3^\mathrm{M}$, where $n_1,n_2,n_3$ are integers. 
In numerical calculation for optical conductivity, we have taken up to $2$th nearest reciprocal
lattices, i.e. $19$ independent reciprocal lattice vectors, which are
the enough number to discuss the low-energy properties of G/hBN system. Thus, the
matrix size for numerical calculations is $38\times 38$.

Figure 2 shows the energy band structure of G/hBN obtained by using continuum model.
When the twist angle gets smaller, the electronic states of G/hBN are
modulated by long periodic moir\'{e} potential. 
At Brillouin zone boundary, energy band gaps appear in valence bands
caused by Bragg scattering where Bloch waves are mixed by 
moir\'{e} periodic potential. 
Owing to the moir\'{e} periodic potential, energy band gap opens about
$20$ meV at hole band side. The position of the energy band gap goes
down
with increasing the twist angle. 
Also, it is noted that a tiny energy band gap with $2$ meV is also
opened at $E=0$. 
\begin{figure*}[ht]
  \begin{center}
    \includegraphics[width=0.8\textwidth]{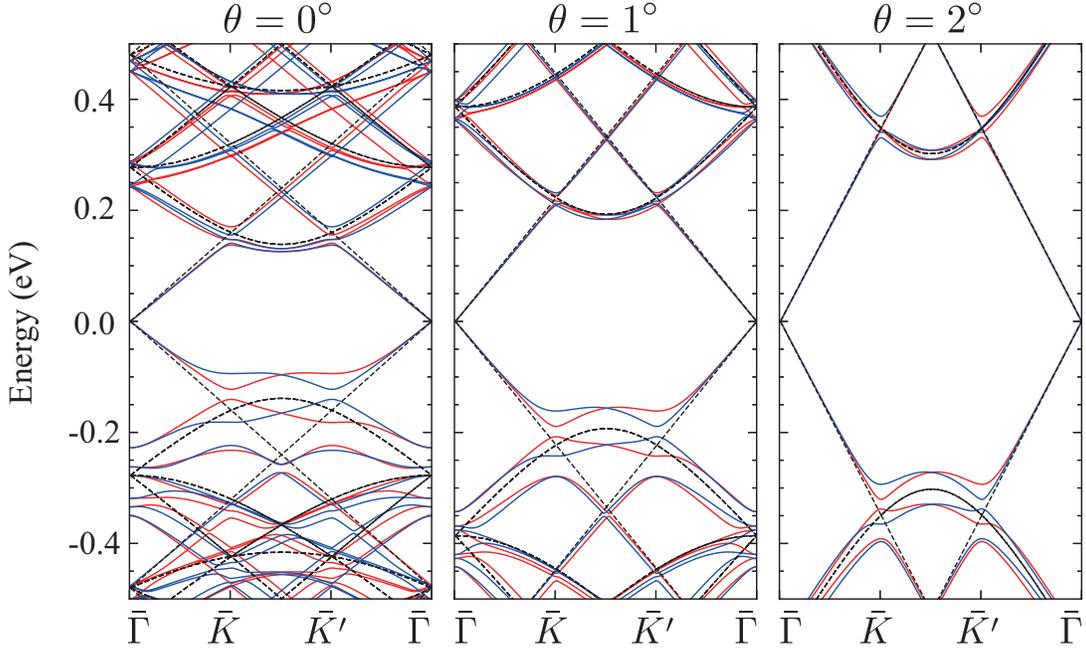}
  \end{center}
  \caption{Energy band structure of G/hBN. Magenta (cyan) bands show the energy bands for $K$ ($K'$) valley and black dashed lines are energy bands of monolayer graphene.}
 \label{fig:fig2}
\end{figure*}

%\subsection{Kubo formula}
We use Kubo formula to calculate optical conductivity of G/hBN.
According to the linear response theory\cite{kubo1957},
the optical conductivity under the light irradiation which has the frequency $\omega$ is given by
\begin{eqnarray}
  \sigma(\omega) &=& \frac{g_s g_ve^2}{i\hbar S} \sum_{i,j}\sum_{\bm{k}}\frac{f(E_{\bm{k}i})-f(E_{\bm{k}j})}{E_{\bm{k}j}-E_{\bm{k}i}}\nonumber\\
  &\quad& \times \frac{|\bm{e}\cdot \bra{\psi_{\bm{k}j}} \nabla_{\bm{k}} H \ket{\psi_{\bm{k}i}}|^2}{E_{\bm{k}j}-E_{\bm{k}i}-\hbar\omega-i\eta},
\end{eqnarray}
where $g_s$ ($g_v$) is spin (valley) degree of freedom, $i,j$ are band indices, $\bm{k}$ is the wavenumber, $E_{\bm{k}i}$ is the eigenenergy for band index $i$, $\psi_{\bm{k}i}$ is the eigenfunction for band index $i$, $S$ is the area of the system, and $f(E)$ is Fermi-Dirac distribution function.
$\eta$ is an infinitesimally small real number.
The vector $\bm{e}$ is the polarization vector of incident light.
Under left-handed circular polarized light (LCP) irradiation, we use
\begin{equation}
  \bm{e}_{\mathrm{LCP}} = \frac{1}{\sqrt{2}}(1, i),
\end{equation}
and under right-handed circular polarized light (RCP) irradiation, we use
\begin{equation}
  \bm{e}_{\mathrm{RCP}} = \frac{1}{\sqrt{2}}(1, -i).
\end{equation}

The matrix element of interband transition $\bra{\psi_{\bm{k}j}}
\nabla_{\bm{k}} H \ket{\psi_{\bm{k}i}}$ is 
related to dipole vector $\bm{D}(\bm{k})$
\begin{align}
\bm{D}(\bm{k})
&\coloneqq  \bra{\psi_{\bm{k}j}} \nabla_{\bm{r}} \ket{\psi_{\bm{k}i}} \\
&= \frac{i}{\hbar}\bra{\psi_{\bm{k}j}} {\bm{p}} \ket{\psi_{\bm{k}i}} \\
&= i\frac{m}{\hbar^2} \bra{\psi_{\bm{k}j}} \nabla_{\bm{k}} H \ket{\psi_{\bm{k}i}},
\label{eq.dipole}
\end{align}
where $\psi_{\bm{k}i}$ is a state of the valence band and
$\psi_{\bm{k}j}$ is a state of the conduction band, respectively.
$\bm{p}=-i\hbar\nabla_{\bm{r}}$ is the momentum operator and 
$m$ is mass of electron.
Dipole vector $\bm{D}(\bm{k})$ is a vector quantity which is composed of complex value.
When the real and imaginary parts of dipole vector are mutually orthogonal,
the circular dichroism under circular polarized light irradiation
occurs\cite{tatsumi2016,ghalamkari2018,akita2020momentum}.  
For monolayer graphene, the behavior of the dynamical conductivity for
massless Dirac electron becomes constant value at the universal
conductivity\cite{ando2002,koshino2013stacking}, 
\begin{equation}
  \sigma_{\mathrm{mono}} = \frac{g_sg_v}{16}\frac{e^2}{\hbar}.
\end{equation}
In followings, we scale the optical conductivity of G/hBN
by using $\sigma_{\mathrm{mono}}$ as
\begin{eqnarray}
  \frac{\sigma(\omega)}{\sigma_{\mathrm{mono}}} &=& \frac{16}{i S^{\mathrm{M}}} \sum_{i,j}\sum_{\bm{k}}\frac{f(E_{\bm{k}i})-f(E_{\bm{k}j})}{E_{\bm{k}j}-E_{\bm{k}i}}\nonumber\\
  &\quad& \times \frac{|\bm{e}\cdot \bra{\psi_{\bm{k}j}} \nabla_{\bm{k}} H \ket{\psi_{\bm{k}i}}|^2}{E_{\bm{k}j}-E_{\bm{k}i}-\hbar\omega-i\eta},
\end{eqnarray}
where $S^{\mathrm{M}}$ is the area of the moir\'{e} periodic unit cell of G/hBN.

Let us consider the optical conductivity of non-doped G/hBN under circular polarized light
irradiation for twist angle $\theta=0^{\circ}$, where
the Fermi energy $E_{\mathrm{F}}$ is $0$ eV.
Figure 3 (a) shows the
optical conductivity of non-doped G/hBN for Hamiltonian with $\xi=+1$. 
The conductivity depends on the direction of rotation of circular
polarized light especially in infrared and terahertz regions. 
In particular, 
for photon energy less than $0.1$ eV, the difference between LCP and RCP
becomes larger.
In this energy region, 
the interband transition from valence to conduction
bands is dominant. 
Figure 3(b) shows the distribution of dipole vectors for 
non-doped case, i.e., $E_{\mathrm{F}} = 0.0$ eV. 
It is clearly seen that 
the real and imaginary parts of
dipole vectors are orthogonal at $\bar{\Gamma}$ point. Thus, 
the circular dichroism is induced by the irradiation of circular
polarized light, and responsible for the electron near $\bar{\Gamma}$ point.
For the plot of dipole dipole vectors of Eq.(\ref{eq.dipole}), we have used the states of
highest valence subband as $\psi_{\bm{k}i}$ below $E_{\mathrm{F}} = 0.0$
eV, and the states of 
lowest conduction subband as $\psi_{\bm{k}j}$ above $E_{\mathrm{F}} =
0.0$.

\begin{figure*}[ht]
  \begin{center}
    \includegraphics[width=0.95\textwidth]{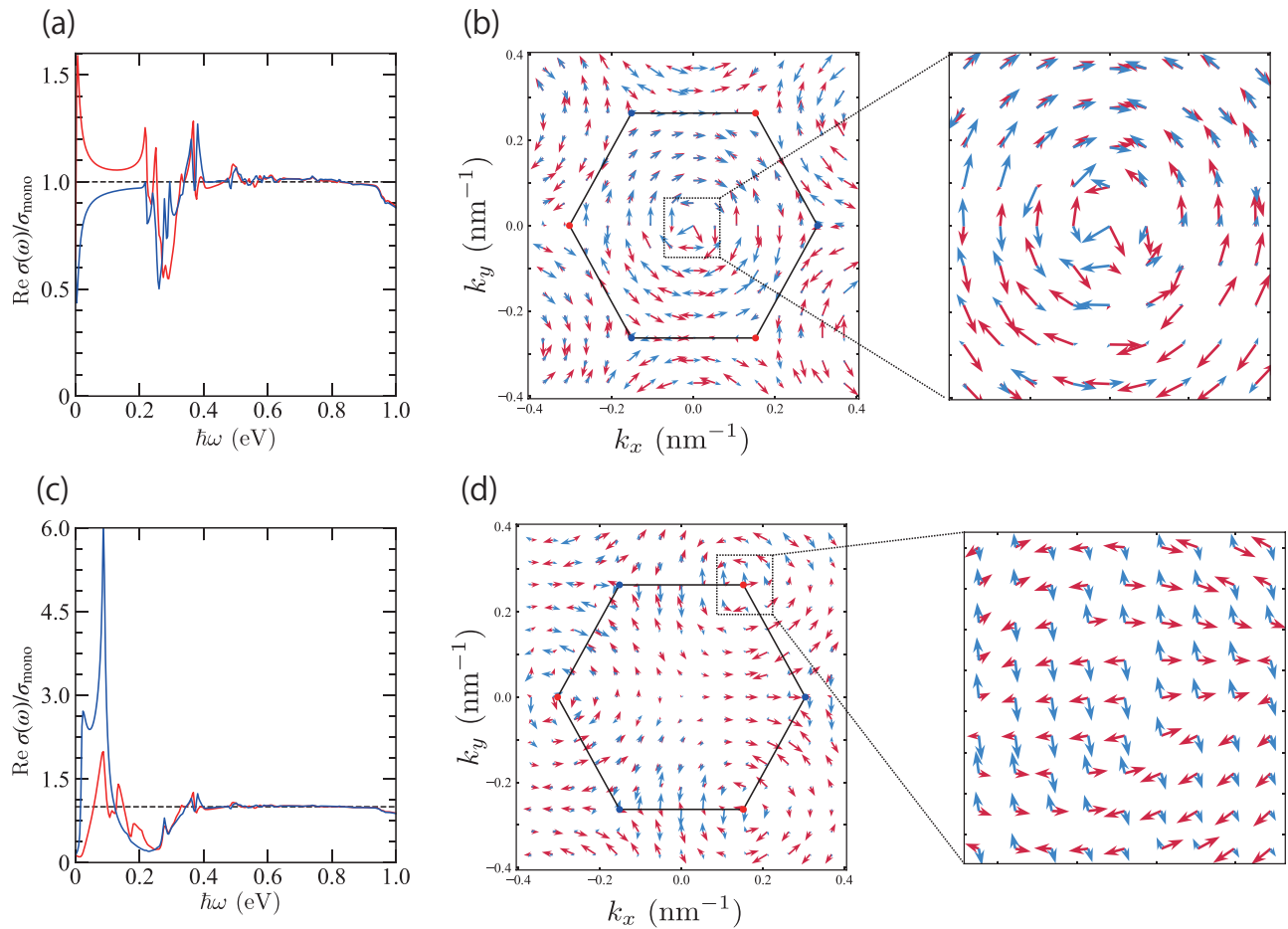}
  \end{center}
  \caption{(a) Optical conductivity of G/hBN for $\xi=1$ and $E_{\mathrm{F}} = 0.0$ eV under circular
 polarized light, where the twist angle $\theta$ is $0^{\circ}$. 
Magenta (cyan) curve is the conductivity for LCP (RCP). 
(b) Distribution of dipole vectors for interband transition near Fermi
 energy $E_{\mathrm{F}} = 0.0$ eV in momentum space. The magenta (cyan)
 arrows are the real (imaginary) part of the dipole vectors. Black hexagon
 is moir\'{e} BZ and magenta (cyan) dots at the hexagon corners are $\bar{K}$ ($\bar{K^\prime}$)
 point. (c) Optical conductivity of G/hBN 
for $E_{\mathrm{F}} = -0.13$ eV, where the twist angle $\theta$ is
 $0^{\circ}$. (d) Distribution of
 dipole vectors for interband transition near Fermi energy
 $E_{\mathrm{F}} = -0.13$ eV in momentum space.} 
 \label{fig:fig3}
\end{figure*}

Next we shall consider the optical conductivity of hole-doped G/hBN. 
Figure~\ref{fig:fig3} (c) shows the optical conductivity with the Fermi
energy $E_{\mathrm{F}}=-0.13$ eV, i.e., hole-doping.
In the region of $\omega < 0.1$ eV, the difference of optical
conductivities between LCP and RCP becomes much larger than the
non-doping case. 
As shown in Fig.~\ref{fig:fig3} (d), 
the real and imaginary parts of dipole vectors are mutually orthogonal
at $\bar{K}$ and $\bar{K}^\prime$ points, which are responsible for the
circular dichroism. 
The results for Hamiltonian with $\xi=-1$ is obtained by
converting LCP (RCP) to RCP (LCP),
because $K^\prime$ states of graphene have opposite chilairty to $K$
states. 
Here, for the plot of dipole dipole vectors of Eq.(\ref{eq.dipole}), we have used the states of
highest occupied subband as $\psi_{\bm{k}i}$ below $E_{\mathrm{F}} = -0.13$
eV, and the states of 
lowest unoccupied subband as $\psi_{\bm{k}j}$ above $E_{\mathrm{F}} =
-0.13$ eV. 

Since the optical conductivity depends on the direction of
rotation of circular polarized light, 
the valley-selective electron excitation is possible in G/hBN.
Here, we define valley polarization for each valley as the difference of
the conductivity between LCP and RCP, 
\begin{equation}
  P(\omega) = \frac{\sigma_{\mathrm{L}} -
   \sigma_{\mathrm{R}}}{\sigma_{\mathrm{L}} + \sigma_{\mathrm{R}}},\
   -1\leq P(\omega) \leq 1,
\end{equation}
where $\sigma_{\mathrm{L}}$ and $\sigma_{\mathrm{R}}$ are the optical
conductivity under LCP and RCP, respectively. 
The positive (negative) value of $P(\omega)$ means that the electron at the valley is easy to excite by LCP (RCP).
Figures 4 (a) and (b) show the photon energy dependence of valley polarization
for (a) non-doping case and (b) hole-doped case, respectively.
In these Figures, the value of $P(\omega)$ is shifted by +1 per
$1^{\circ}$ of twist angle.
The black, magenta and cyan curves indicate the case of twist angle
$\theta=0, 1, 2^{\circ}$. 
For non-doping case, the value of valley polarization less than $0.1$ eV
has maximum value for $\theta=0^{\circ}$ and it is decreasing with
increase of the twist angle. 
These results indicate that the electronic properties of G/hBN approaches
to the electronic states of monolayer graphene with increase of the
twist angle. 
On the other hand, the peak of valley polarization less than $0.1$ eV
for hole-doped case remains even though the twist angle becomes larger. 
The common peak of valley polarization comes from interband transition
across the Fermi energy $E_{\mathrm{F}}$ at $\bar{K}$ point. 
\begin{figure*}[ht]
  \begin{center}
    \includegraphics[width=0.80\textwidth]{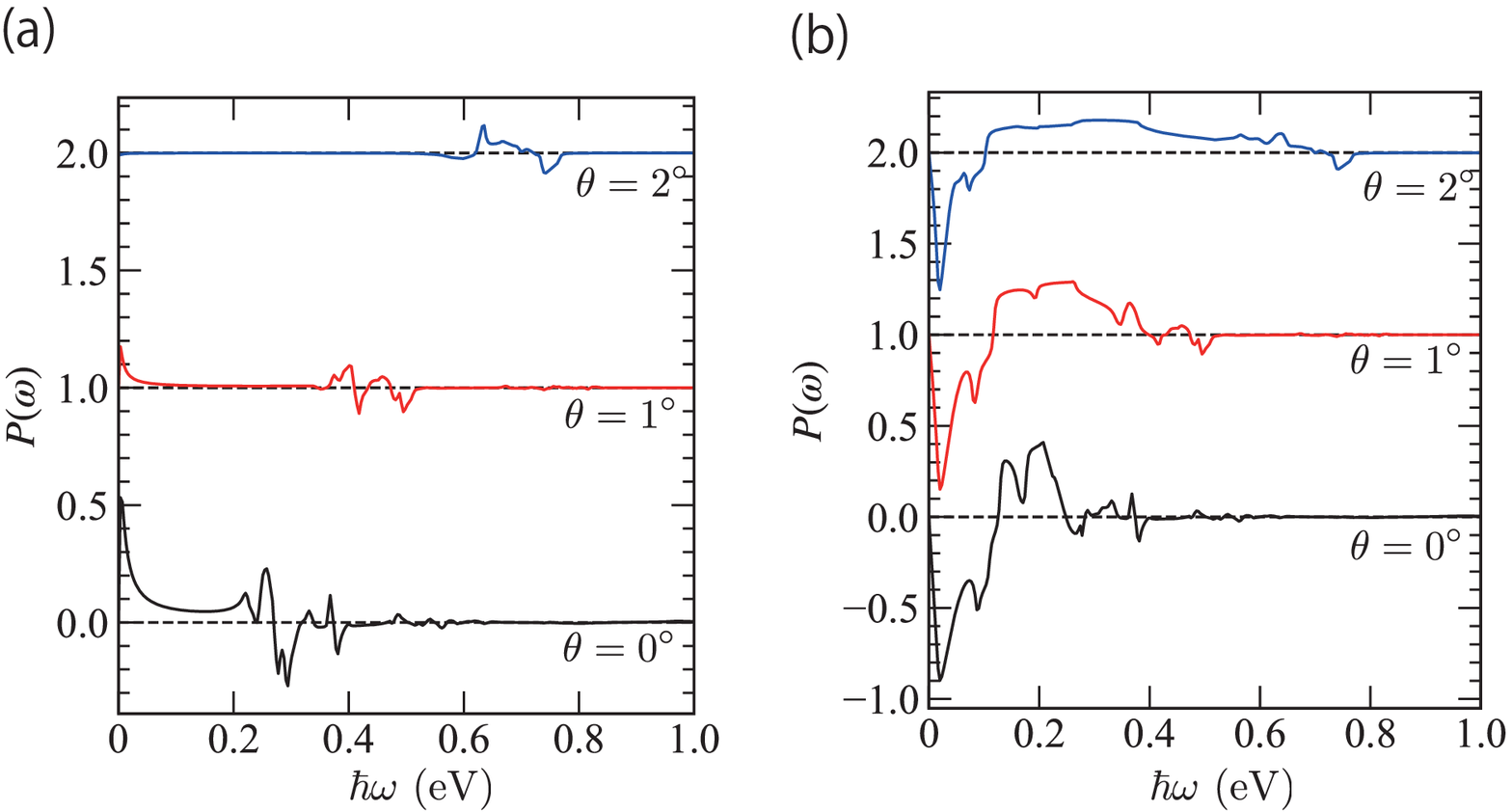}
  \end{center}
  \caption{(a) Photon energy dependence of valley polarization
 ($E_{\mathrm{F}} = 0$ for several different twist angles.
 eV). Black dashed lines indicate $P(\omega)=0$ line.
 (b) Photon energy dependence of valley polarization for hole-doped
 case, where the Fermi energy is set in the bandgap of the valence band at $\bar{K}$
 point. Fermi energies are set $E_{\mathrm{F}} = -0.13$ eV for
 $\theta=0^{\circ}$, $E_{\mathrm{F}} = -0.20$ eV for $\theta=1^{\circ}$,
 and $E_{\mathrm{F}} = -0.33$ eV for $\theta=2^{\circ}$.} 
 \label{fig:fig4}
\end{figure*}

In this paper, we have numerically calculated the optical conductivity of G/hBN
under circular polarized light irradiation. 
The optical conductivity of G/hBN changes their behavior depending on the
direction of rotation of circular polarized light due to broken
spatial inversion symmetry of the system. 
By analyzing the dipole vectors, we have confirmed that the real and
imaginary parts of dipole vectors are mutually orthogonal.
Thus, we can explain that the electron for each valley can be
selectively excited by using a certain direction of circular polarized
light irradiation. 
These results indicate that G/hBN can cause valley polarization by circular
polarized light irradiation. 
Also, we confirmed that the value of valley polarization of G/hBN
depends on not only twist angle but also Fermi energy. 
For nondoping case, the peak of valley polarization for photon energy
less than $0.1$ eV is vanished as larger twist angle. 
On the other hand, for hole-doped case, it remains even though twist angle becomes larger.
These results will serve to design the valleytronics devices using G/hBN.

This work was supported by JSPS KAKENHI (No. JP21H01019, No. 22H05473 and No. JP18H01154) and JST CREST (No. JPMJCR19T1).
%%%reference
\bibliography{reference}
\end{document}